\documentclass[aps,prb,amsmath,amssymb,amsfont,showpacs,superscriptaddress,reprint]{revtex4-1}
\usepackage{graphicx}
\usepackage{color}
\definecolor{dgray}{gray}{0.3}
\definecolor{mgray}{gray}{0.5}
\definecolor{lgray}{gray}{0.7}
\definecolor{dgreen}{rgb}{0,0.7,0}
\usepackage{times}
\usepackage{natbib}
\usepackage{subfigure}

\begin{document}

\newcommand{\dd}[1]{\mathrm{d}#1}
\newcommand{\kb}{k_\text{B}}
\newcommand{\Td}{T_\text{D}}
\newcommand{\comment}[2]{#2}
\newcommand{\todo}[1]{{\color{red}#1}}

\title{The empirical Monod-Beuneu relation of spin-relaxation revisited for elemental metals}

\author{L. Szolnoki}
\affiliation{Institute of Physics of Complex Matter, FBS Swiss Federal Institute of Technology (EPFL), CH-1015 Lausanne, Switzerland}
\affiliation{Department of Physics, Budapest University of Technology and Economics and Condensed Matter Research Group of the Hungarian Academy of Sciences, Budafoki \'{u}t 8, H-1111 Budapest, Hungary}

\author{A. Kiss}
\affiliation{Wigner Research Centre for Physics of the Hungarian Academy of Sciences, Budapest, Hungary}
\affiliation{BME-MTA Exotic Quantum Phases Research Group, Budapest University of Technology and Economics, Budapest, Hungary}

\author{L. Forr\'o}
\affiliation{Institute of Physics of Complex Matter, FBS Swiss Federal Institute of Technology (EPFL), CH-1015 Lausanne, Switzerland}

\author{F. Simon}
\email[Corresponding author: ]{ferenc.simon@univie.ac.at}
\affiliation{Department of Physics, Budapest University of Technology and Economics and Condensed Matter Research Group of the Hungarian Academy of Sciences, Budafoki \'{u}t 8, H-1111 Budapest, Hungary}


\date{\today}

\begin{abstract}
Monod and Beuneu [Monod and Beuneu, Phys. Rev. B \textbf{19}, 911 (1979)] established the validity of the Elliott-Yafet theory for elemental metals through correlating the experimental electron spin resonance line-width with the so-called spin-orbit admixture coefficients and the momentum-relaxation theory. The spin-orbit admixture coefficients data were based on atomic spin-orbit splitting. We highlight two shortcomings of the previous description: i) the momentum-relaxation involves the Debye temperature and the electron-phonon coupling whose variation among the elemental metals was neglected, ii) the Elliott-Yafet theory involves matrix elements of the spin-orbit coupling (SOC), which are however not identical to the SOC induced energy splitting of the atomic levels, even though the two have similar magnitudes. {\color{black}We obtain the empirical spin-orbit admixture parameters for the alkali metals by considering the proper description of the momentum relaxation theory. \textcolor{black}{In addition,} we present a model calculation which highlights the difference between the SOC matrix element and energy splitting.}
\end{abstract}

\maketitle

\section{Introduction}

Information storage and processing using spins, referred to as spintronics \cite{FabianRMP}, is an actively studied subject \cite{WuReview}. The interest has been renewed by the prospect of using graphene for spintronics although the results are as yet controversial \cite{HuertasPRB2006,FabianPRB2009a,FabianPRB2009b,CastroNetoGrapheneSO,DoraEPL2010,WuNJP2012,CastroNetoGuineaPRL2012}.

Spintronics exploits that spin-relaxation time, $\tau_{\text{s}}$, exceeds the momentum-relaxation time, $\tau$, by several orders of magnitude. $\tau_{\text{s}}$ gives the characteristic timescale on which a non-equilibrium spin-ensemble, either induced by electron spin resonance \cite{KipKittelPR1952} or by a spin-polarized current \cite{JohnsonSilsbeePRB1988,JedemaNat2002}, decays to the equilibrium. It is thus the central parameter which characterizes the effectiveness of spin-transport and eventually the utility of spintronics.

In metals with inversion symmetry, the mechanism of spin-relaxation is described by the Elliott-Yafet (EY) theory \cite{Elliott,yafet1963g}. In the absence of spin-orbit coupling (SOC), there is no relaxation between the spin-up/down states. However, SOC induces spin mixing and the resulting admixed states read:

{
\begin{eqnarray}
{\mid\! \widetilde{+} \rangle}_\mathbf{k} &=& \left[a_\mathbf{k}\left(\mathbf{r}\right)\mid\! + \rangle + b_\mathbf{k}\left(\mathbf{r}\right) \mid\! - \rangle\right]e^{i\mathbf{kr}}, \label{eq-st1}\\
{\mid\! \widetilde{-} \rangle}_\mathbf{k} &=& \left[a^{\ast}_{-\mathbf{k}}\left(\mathbf{r}\right)\mid\! - \rangle - b^{\ast}_{-\mathbf{k}}\left(\mathbf{r}\right) \mid\! + \rangle\right]e^{i\mathbf{kr}},\label{eq-st2}
\end{eqnarray}
}

\noindent where ${\mid\! + \rangle}$ and ${\mid\! - \rangle}$ are the pure spin states and  ${\mid\! \widetilde{+} \rangle}_\mathbf{k}$, ${\mid\! \widetilde{-} \rangle}_\mathbf{k}$ are the perturbed Bloch states. The admixture strength is given by the so-called spin-orbit admixture coefficient (SOAC), which in the first order of the SOC is:
{
$\frac{|b_\mathbf{k}|}{|a_\mathbf{k}|}\propto \frac{L}{\Delta E}$,
}
where $L$ is the matrix element \cite{Note}
of the SOC for the conduction and the near lying band with an energy separation of $\Delta E$. We note that for metals with inversion symmetry, the admixed spin-up/down states of the conduction band remain degenerate in the absence of magnetic field due to the time reversal symmetry (or Kramers' theorem).

Elliott showed \cite{Elliott} that the usual momentum-scattering induces spin transitions for the admixed states, i.e. a spin-relaxation, whose magnitude is:

\begin{gather}
\frac{1}{\tau_{\text{s}}}=\alpha_1 \left( \frac{L}{\Delta E}\right)^2 \frac{1}{\tau},
\label{elliottrel1}
\end{gather}

\noindent where $\alpha_1$ is a band structure dependent constant near unity.

Elliott further showed that the magnetic energy of the admixed states is different from that of the pure spin-states, i.e. there is a shift in the electron gyromagnetic factor, or $g$-factor:

\begin{gather}
\Delta g=g-g_0=\alpha_2 \frac{L}{\Delta E},
\label{elliottrel2}
\end{gather}

\noindent where $g_0 \approx 2.0023$ is the free electron $g$-factor, $\alpha_{2}$ is another band structure dependent constant near unity. Eqs. \eqref{elliottrel1} and \eqref{elliottrel2} result in the so-called Elliott relation

\begin{equation}
\frac1{\tau_{\text{s}}}=\frac{\alpha_1}{\alpha_2^2}\frac{\Delta g^2}{\tau},
\label{elliottrel3}
\end{equation}

\noindent which links three empirical measurables; $\tau_{\text{s}}$, $\tau$, and $\Delta g$. In practice, the spin-relaxation time is obtained for metals from conduction electron spin resonance (CESR) measurements \cite{FeherKip}. This yields $\tau_{\text{s}}$ directly from the homogeneous ESR line-width, $\Delta B$ through $\tau_{\text{s}} = \left(\gamma\Delta B\right)^{-1}$, where $\gamma/2\pi= 28.0 \,\text{GHz/T}$ is the electron gyromagnetic ratio. The CESR resonance line position yields the $g$-factor shift.

Although, the original theory of Elliott \cite{Elliott} involves the momentum-scattering time, $\tau$, the transport momentum-scattering time, $\tau_{\text{tr}}$ is more readily obtained from the specific resistivity, $\rho$ through: $\rho^{-1}=\epsilon_0 \omega_{\text{pl}}^2 \tau_{\text{tr}}$, where $\epsilon_0$ is the vacuum permittivity, $\omega_{\text{pl}}$ is the plasma frequency. The two momentum-scattering times differ in a constant at high temperature but have a characteristically different temperature dependence at low $T$: $\tau \propto T^{-3}$ and $\tau_{\text{tr}} \propto T^{-5}$ {\color{black}(for scattering due to phonons)}. Yafet showed that the low temperature spin-relaxation time also follows a $T^{-5}$ law \cite{yafet1963g}. This allows to summarize the Elliott-Yafet relation as:

\begin{equation}
\Delta B=\frac{\alpha_1}{\alpha_2^2}\Delta g^2 \epsilon_0 \omega_{\text{pl}}^2  \rho.
\label{elliottrel_with_rho}
\end{equation}

Monod and Beuneu contributed to the field with two seminal papers \cite{BeuneuMonodPRB1978,MonodBeuneuPRB1979}:
in Ref.~\onlinecite{BeuneuMonodPRB1978} they tested the Elliott-relation by collecting $\Delta B$ and $\Delta g$ data for elemental metals. They found that the Elliott-relation is valid with $\frac{\alpha_1}{\alpha_2^2} \approx 10$ for alkali metals (except for Li) and for monovalent transition metals (Cu, Ag, and Au). It is interesting to note that the validity of the Elliott-relation has since been confirmed for alkali fullerides \cite{PetitPRB1997} and intercalated graphite \cite{FabianPRB2012}. Deviations from the Elliott-relation for polyvalent metals (such as Mg and Al) was explained by Fabian and Sarma by considering the unique details of the band structure where the SOC is enhanced, which is known as the "hot-spot" model \cite{FabianPRL1998,FabianPRL1999}.

\begin{figure}[htp]
\begin{center}
\includegraphics[scale=.5]{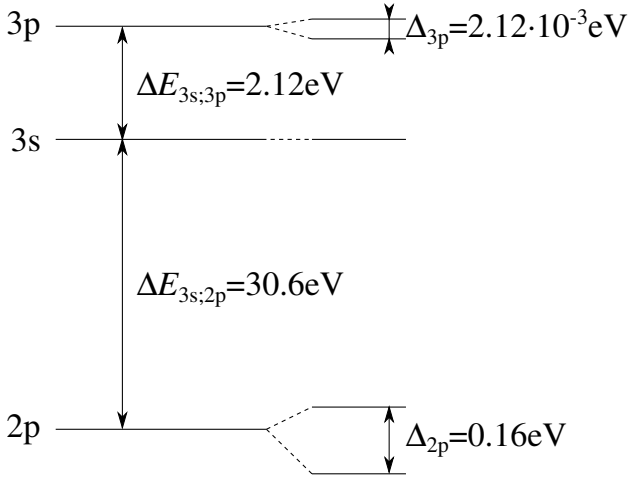}
\caption{The level scheme (not to scale) which is relevant for the spin-orbit admixture in Na. Note that $\left(\Delta_{\text{3p}}/\Delta E_{\text{3s;3p}}\right)<\left(\Delta_{\text{2p}}/\Delta E_{\text{3s;2p}}\right)$, the latter therefore dominates the SOAC.}
\label{LevelScheme}
\end{center}
\end{figure}

In their second seminal paper (Ref. \onlinecite{MonodBeuneuPRB1979}), Monod and Beuneu attempted
to correlate the spin-relaxation data with estimated spin-orbit admixture constants. The energy splitting of a relevant atomic state due to SOC was used as an estimate for the matrix element of the SOC between the conduction and a near lying state. E.g. for Na, {\color{black}definition of the relevant quantities is given in Fig. \ref{LevelScheme}} and the conduction band is the $\text{3s}$ state; the SOAC is either the $\Delta_{\text{3p}}/\Delta E_{\text{3s;3p}}$ or $\Delta_{\text{2p}}/\Delta E_{\text{3s;2p}}$, whichever of the two ratios is the greater. For Na, it is the $\Delta_{\text{2p}}/\Delta E_{\text{3s;2p}}$ ratio and the situation is depicted in Fig. \ref{LevelScheme}. Monod and Beuneu found that the ESR line-width data, when normalized by the larger of the two possible ratios squared, $\Delta B \cdot \left(\frac{\Delta E}{L}\right)^2$, falls on the \emph{same} universal Gr\"{u}neisen function for the alkali atoms (Na, K, Rb, and Cs) and for the monovalent transition metals (Cu, Ag and Au) as a function of the normalized temperature $T/T_{\text{D}}$ ($T_{\text{D}}$ is the Debye temperature). Much as Ref.~\onlinecite{MonodBeuneuPRB1979} became a standard for our understanding of the spin-relaxation in elemental metals, it has some shortcomings and widespread misinterpretations in the literature which motivates the present revision.

First, the transport momentum-relaxation time scales with the transport electron-phonon coupling, $\lambda_{\text{tr}}$, and the Debye temperature, $\Td$, which was neglected in Ref. \onlinecite{MonodBeuneuPRB1979}. Second, it is not immediately clear why the SOC induced atomic \emph{energy splittings} should be identical to the spin-orbit \emph{matrix elements}, even though one expects similar orders of magnitude. This uncertainty led to a confusion concerning what is meant by the SOC strength (e.g. Refs.~\onlinecite{BMref5, BMref13, BMref17, BMref18, BMref23, BMref31, BMref37, BMref38, BMref41}). When investigated in detail, one finds that the agreement between the scaled ESR line-width and the "universal" Gr\"{u}neisen function is a result of the neglected $\Td$ and $\lambda_{\text{tr}}$ dependence. We note that the first hint that the atomic picture is not sufficient to explain the spin-relaxation properties came from the above mentioned works of Fabian and Sarma \cite{FabianPRL1998,FabianPRL1999} who showed that band-structure effects play an important role in aluminium and in other polyvalent metals.

Herein, we show that in Ref.~\onlinecite{MonodBeuneuPRB1979} the variation of the transport electron-phonon coupling constant and $T_\text{D}$ among the different metals was neglected, which however affects the value of $\tau_{\text{tr}}$. We show that the agreement between the scaled ESR line-width and the "universal" Gr\"{u}neisen function, which was found in Ref.~\onlinecite{MonodBeuneuPRB1979} is a result of the neglected $\Td$ and $\lambda_{\text{tr}}$ dependencies. We present an analysis to provide the \emph{empirical} spin-orbit admixture coefficients, which could serve as an input for future first principles based calculations. We also show that while the atomic spin-orbit splitting energies have the same order of magnitude as the matrix elements of the SOC between adjacent bands but they are not identical. {\color{black}We provide a model calculation involving s and p states with spin-orbit coupling to explicitly show that the atomic SOC induced energy splitting is not identical to the SOC matrix element, the latter being sensitive to the s-p hybridization, i.e. for the details of the band-structure.}

\section{Results and discussion}

\subsection{The spin-orbit admixture parameters}

In Ref.~\onlinecite{MonodBeuneuPRB1979}, Monod and Beuneu investigated the scaling of the normalized ESR line-width with the transport momentum-relaxation time, $\tau_{\text{tr}}$, and found that the normalized ESR line-width data falls on a universal Gr\"{u}neisen function\cite{mott1936theory}:

\begin{equation}
\begin{aligned}
\Delta B \cdot \left(\frac{\Delta E}{L}\right)^2 &=\text{const} \cdot \frac{T}{T_\text{D}} \cdot G_{\text{MB}}\left(\frac{T_\text{D}}{T}\right), \\
\text{where}\quad G_\text{MB}(x)&=4x^{-4}\left[5\int_0^x\frac{z^4 \dd{z}}{e^z-1}-\frac{x^5}{e^x-1}\right],
\end{aligned}%
\label{grun_funcBM}
\end{equation}

\noindent where the constant was considered to be metal independent. The $L/\Delta E$ SOAC data were based on atomic spectra and were taken from Ref.~\onlinecite{yafet1963g}. The Gr\"{u}neisen function, $G_{\text{MB}}$, used by Monod and Beuneu was taken from Ref.~\onlinecite{mott1936theory}. The original paper, Ref.~\onlinecite{MonodBeuneuPRB1979}, did not explicitly mention the normalization with $T_\text{D}$. However since a single, "universal" Gr\"{u}neisen function was argued to represent well the data \cite{MonodBeuneuPRB1979}, this presentation implies the $T_{\text{D}}^{-1}$ factor. This, as we show below, makes the value of the SOAC uncertain. The role of the spin-orbit coupling admixture is discussed further below and here we first focus on the parameters of the transport momentum-scattering theory.

The contemporary description of the transport momentum-relaxation for alkali metals within the Debye-model assuming zero residual scattering reads \cite{poole2000handbook}:

\begin{equation}
\begin{aligned}
\frac1{\tau_{\text{tr}}}&=\frac{2\pi \kb }{\hbar}\lambda_{\text{tr}} T \cdot G\left(\frac{T}{\Td }\right), \\
\text{where}\quad G(x)&=\int_0^1\dd{u}\frac{u^5}{x^2\sinh^2\left(u/(2x)\right)},
\end{aligned}%
\label{grun_func_contemporary}
\end{equation}

\noindent where $k_{\text{B}}$ and $\hbar$ are the Boltzmann and Planck constants, respectively and $\lambda_{\text{tr}}$ is the transport electron-phonon coupling constant. The two forms of the Gr\"{u}neisen function, $G(x)$ and $G_{\text{MB}}(1/x)$, in Eqs. \eqref{grun_funcBM} and \eqref{grun_func_contemporary} are equivalent.

Eq. \eqref{grun_func_contemporary} when substituted into Eq.~\eqref{elliottrel1} reads for the normalized ESR line-width:

\begin{equation}
\begin{aligned}
\Delta B\cdot \left(\frac{\Delta E}{L}\right)^2&=\alpha_1 \frac{2\pi \kb }{\gamma\hbar} \lambda_{\text{tr}} T\cdot G\left(\frac{T}{\Td }\right). \\
\label{ESR_linewidth_rewritten}
\end{aligned}
\end{equation}

Clearly, an uncertainty remains due to the parameter $\alpha_1$, which is however supposed to be around unity and the same for all alkali metals \cite{Elliott}. Eq. \eqref{ESR_linewidth_rewritten} allows to introduce a universal function:

\begin{equation}
\begin{aligned}
F(x)=\frac{2\pi \kb }{\gamma\hbar} x G\left(x\right), \\
\label{F_introduced}
\end{aligned}
\end{equation}

\noindent which yields the final result of

\begin{equation}
\begin{aligned}
\Delta B\cdot \left(\frac{\Delta E}{L}\right)^2&=\alpha_1 \Td \lambda_{\text{tr}} F\left(\frac{T}{\Td }\right).\\
\label{ESR_linewidth_rewritten3}
\end{aligned}
\end{equation}

The left-hand side of Eq. \eqref{ESR_linewidth_rewritten3} is proportional to $\alpha_1$, $\Td $ and $\lambda_{\text{tr}}$. However Monod and Beuneu plotted the measured ESR line-widths while neglecting the variation of $\Td  \cdot \lambda_{\text{tr}}$ among the alkali metals, even though it can amount to a factor 4.

\begin{table}
\caption{\label{tab:lambdas}The electron-phonon coupling constants from Ref. \onlinecite{AllenLambdas} and Debye-temperatures from Ref. \onlinecite{kittel1996introduction} of alkali elements. We also give the $(L/\Delta E)^2$ values from Ref. \onlinecite{MonodBeuneuPRB1979} (in the original notation $(\lambda/\Delta E)^2$ ). The fitted values of $(L/\Delta E)^2$ are determined herein.}
\begin{ruledtabular}
\begin{tabular}{ccccc}
Alkali element & \multicolumn{1}{c}{$\lambda_{\text{tr}}$} & $\Td\,\text{[K]}$ & atomic $(L/\Delta E)^2$ & fitted $(L/\Delta E)^2 $\\
\hline
Na	& 0.14 & 158 & $2.73 \cdot 10^{-5}$ & $3.81 \cdot 10^{-6}$ \\
K	& 0.11 & 91 & $2.06 \cdot 10^{-4}$ &  $8.99 \cdot 10^{-5}$ \\
Rb	& 0.15 & 56 & $3.16 \cdot 10^{-3}$ &  $2.96 \cdot 10^{-3}$ \\
Cs	& 0.16 & 38 & $1.91 \cdot 10^{-2}$ &  $3.08 \cdot 10^{-2}$ \\
\end{tabular}
\end{ruledtabular}
\end{table}

In Table \ref{tab:lambdas}., we give values of $\lambda_{\text{tr}}$ and $\Td$ for the four alkali metals. We also give the SOAC values as used by Monod and Beuneu for the scaling. We proceed with the analysis of the available data by using the values of $\lambda_{\text{tr}}$ and $\Td$ given in Table \ref{tab:lambdas}. The $\Delta B\cdot \left(\frac{\Delta E}{L}\right)^2$ data is taken from Ref. \onlinecite{MonodBeuneuPRB1979}.

\begin{figure}[htp]
\begin{center}
\includegraphics[scale=.32]{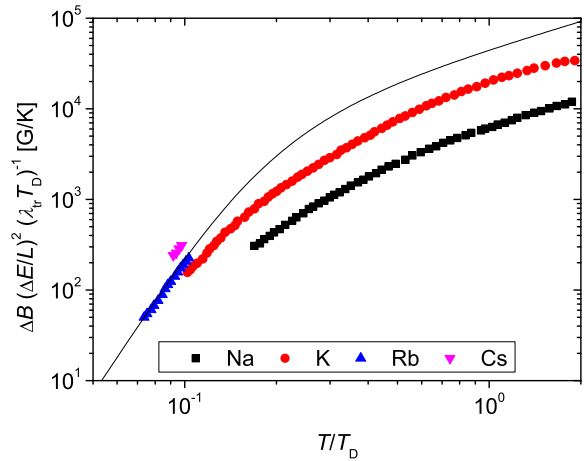}
\caption{The experimental $\Delta B \cdot \left({\Delta E}/L\right)^2 /\lambda_{\text{tr}} \Td$ plotted against $T/\Td$. It is important to note that the atomic values of $\left({\Delta E}/L\right)^2$ are used herein for the scaling (such as it was done by Monod and Beuneu). Solid curve shows the universal $F(x)$ function after Eq. \eqref{F_introduced}. Note that the line-width data do not fall on the same universal curve.}
\label{gr04}
\end{center}
\end{figure}

In Fig.~\ref{gr04}., we show $\Delta B\cdot \left(\frac{\Delta E}{L}\right)^2/\Td \lambda_{\text{tr}}$ versus $T/\Td$. The universal $F(x)$ function from Eq. \eqref{F_introduced} is also shown. Clearly, the normalized line-width data do not fall on the same curve when the variation of $\lambda_{\text{tr}}$ and $\Td$ among the four alkali metals is taken into account. {\color{black}This means that the atomic SOC induced energy splitting per the energy difference between the adjacent states do not approximate well the real SOAC values.} Accidentally, the data for {\color{black}Rb} lies well on the plot indicating that then the proper SOAC value is well approximated by the atomic one.

\begin{figure}[htp]
\begin{center}
\includegraphics[scale=.3]{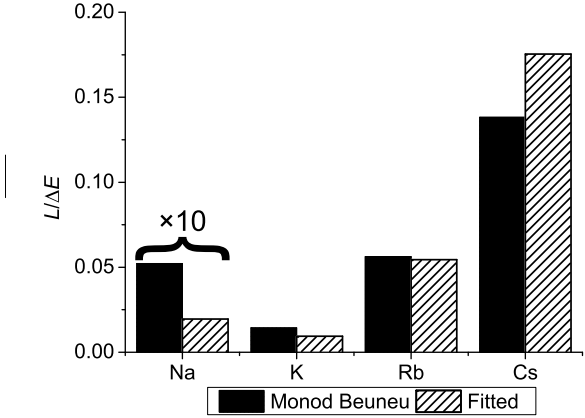}
\caption{Comparison of the herein determined spin-orbit admixture coefficients and the values used by Monod and Beuneu in Ref. \onlinecite{MonodBeuneuPRB1979}. The values for Na are multiplied by 10 for better visibility. {\color{black}Note the agreement for Rb between the present values and those determined previously.}}
\label{lperd}
\end{center}
\end{figure}

 Once the relevance of $\Td$ and $\lambda_{\text{tr}}$ is recognized, we use the experimental data to determine the experimental SOAC. In Fig. \ref{lperd}., we show the SOAC values which are determined herein and those considered by Monod and Beuneu in Ref. \onlinecite{MonodBeuneuPRB1979}. We observe a non-negligible difference between the values used previously and those which are obtained considering the role of $\lambda_{\text{tr}}$ and $\Td$. The present empirical values could be used as input for improved first principles calculations, which consider the band structure of these elements including spin-orbit coupling. Naturally, such calculations were unavailable at the time of Ref. \onlinecite{MonodBeuneuPRB1979}, therefore our refinement of the values do not detract from the merit of the original work which highlighted the role of the atomic spin-orbit coupling.

\subsection{The matrix element of the spin-orbit coupling}

As mentioned above, Monod and Beuneu \cite{MonodBeuneuPRB1979} estimated the
spin-orbit admixture coefficients, $L/\Delta E$, using values based on atomic ones: for $L$, the atomic SOC induced energy splitting of a p orbital (adjacent to an s orbital based conduction band) and for $\Delta E$ the corresponding energy separation was used. While the energy separation between atomic orbitals is a good approximation for band-band separations (given that usual band-widths are an order of magnitude smaller than energy separations in alkali metals), $L$ is a \emph{matrix element} between neighboring s and p orbitals in the Elliott theory and not the \emph{energy splitting} for a p orbital. It is therefore not straightforward why the energy splitting should equal the matrix element of the SOC between the s and p orbitals.

{
The Elliott-Yafet theory involves the matrix elements of the SOC Hamiltonian, which reads for a radial symmetry of the interaction as
\begin{equation}
\begin{aligned}
H_{\text{SO}} &= \frac{\hbar^2}{2 m_0^2c^2}\frac1r\frac{\partial V}{\partial r}  \mathbf{L}\cdot\mathbf{S} = \lambda(r) \mathbf{L}\cdot\mathbf{S},
\end{aligned}
\end{equation}
where $m_0$ is the free electron mass.
We denote the matrix elements of the SOC by $L_{n;n'}$ between the conduction band indexed by $n$ and an adjacent one with $n'$, and the corresponding energy separation between the bands with $\Delta E_{n;n'}$.
The spin-relaxation is dominated by that neighboring band for which the $L_{n;n'}/\Delta E_{n;n'}$ ratio is larger. E.g. for alkali metals, the conduction band is based on the $n$, $\text{s}$ orbital and the dominant spin-orbit state turns out to be the $n-1$, $\text{p}$ state (see Fig.~\ref{LevelScheme}).
}

\begin{figure}[htp]
\begin{center}
\includegraphics[scale=.3]{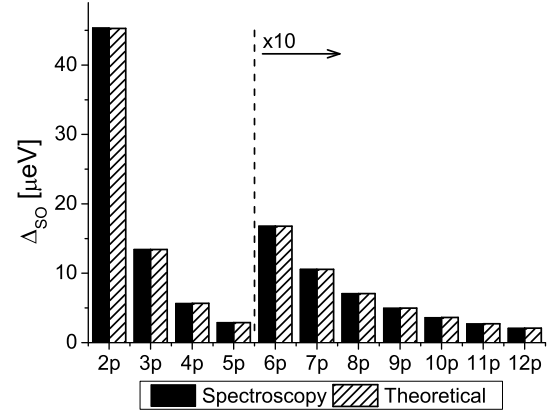}
\caption{Comparison between the SOC splittings for hydrogen calculated according to Eq.~(\ref{HydrogenLikeAtomSOCSplitting}) and the experimental, spectroscopy based values from Ref. \onlinecite{kramida2010critical}.}
\label{HydrogenModellSplittingvsTheory}
\end{center}
\end{figure}

{
In the presence of the SO interaction, the sixfold degenerate atomic p state splits in accord with $j=3/2$ and $j=1/2$, where $j$ is the  total angular momentum which becomes a good quantum number instead of $l$ and $s$.
The SO matrix elements are given for the hydrogen as
\begin{eqnarray}
L_{j=3/2} = \frac{1}{2} l \lambda, \,\,\,\,\, L_{j=1/2} = -\frac{1}{2} (l +1 ) \lambda
\end{eqnarray}
with
\begin{eqnarray}
\lambda = \int_{0}^{\infty} R_{n, l}^2(r)\lambda (r)r^2 \dd r,
\end{eqnarray}
where $R_{n,l}(r) $ denotes the radial component of the hydrogen wave functions. Thus, the energy splitting of the p state is expressed as
\begin{eqnarray}
\Delta_\text{p} = L_{j=3/2} - L_{j=1/2} = 1/2(2l+1)\lambda.
\label{HydrogenLikeAtomSOCSplitting}
\end{eqnarray}
}
In Fig.~\ref{HydrogenModellSplittingvsTheory}., we show the comparison between the energy splittings $\Delta_\text{p}$ with different $n$ and the experimental data available from Ref.~\onlinecite{kramida2010critical}. The two sets of data match within $0.2\,\%$, which demonstrates that the calculation works accurately for hydrogen.

{
{\color{black}Monod and Beuneu used the SOC induced \emph{energy splitting} parameters for the SO matrix elements involved in the Elliott-Yafet theory.} We demonstrate herein that the two quantities are not equal in general, i.e. $L_{n;n'} \neq \Delta_{n'}(=\Delta_{\text{SO}})$
even though they both originate from the SOC.}

{
As the simplest model to discuss spin-relaxation in alkali metals such as Na shown in Fig.~\ref{LevelScheme}, we consider electrons moving on a simple cubic lattice with an s and a p state at each site\cite{yanaseKB2011}.
The Hamiltonian of this model is given as
\begin{eqnarray}
{\cal H} &=& {\cal H}_{0} +  {\cal H}_{\text{SO}}; \label{eq-ham11}\\
{\cal H}_{0} &=& {\cal H}_{\text{kin}} + {\cal H}_{\text{hyb}} + {\cal H}_{\text{s}} + {\cal H}_{\text{p}}
\nonumber\\
&=& \sum_{i, \boldsymbol{\delta}} \sum_{\sigma} \sum_{m=\text{s},x,y,z} t_{m} c^{\dag}_{i,m \sigma} c_{i+\boldsymbol{\delta}, m \sigma}
\nonumber\\
&+& \sum_{i, \boldsymbol{\delta}} \sum_{\sigma}  \sum_{m=x,y,z} v_{\text{s}m,\boldsymbol{\delta}} \left( c^{\dag}_{i,\text{s}\sigma} c_{i+ \boldsymbol{\delta}, m\sigma} + {\rm h.c.} \right)\nonumber\\
&+&
 E_\text{s} \sum_{i \, \sigma}   c^{\dag}_{i,\text{s} \sigma} c_{i, \text{s} \sigma}
 +  E_\text{p} \sum_{i \, \sigma} \sum_{m=x,y,z}   c^{\dag}_{i,m \sigma} c_{i, m \sigma}
 , \label{eq-ham12} \\
{\cal H}_{\text{SO}} &=&  \lambda \sum_{i} \boldsymbol{L}_{i} \cdot \boldsymbol{S}_{i},
\label{eq-ham13}
\end{eqnarray}
where we regard the spin-orbit interaction ${\cal H}_{\text{SO}}$ as perturbation in addition to the principal part, ${\cal H}_{0}$, that includes the kinetic energy \textcolor{black}{with the hopping parameters $t_{m}$}, the s-p mixing \textcolor{black}{described by the hybridization parameters $v_{\text{s}m,\boldsymbol{\delta}}$}, and the s and p state on-site energies \textcolor{black}{$E_{s}$ and $E_{p}$, respectively}.
The operator $c^{\dag}_{i,m \sigma}$ creates an electron with spin $\sigma$ and orbital $m$ at the lattice site $i$,
and $v_{\text{s}m,\boldsymbol{\delta}} = v_{\text{s}m}\mathbf{e}_m \boldsymbol{\delta}$ with $\boldsymbol{\delta}$ being a vector that points to a neighboring site \textcolor{black}{and $\mathbf{e}_m$ is a unit-vector parallel to the $m$ axis.}
After Fourier transformation, we obtain the band energies from the hopping and the hybridization terms as
\begin{eqnarray}
\varepsilon_{m}(\boldsymbol{k}) &=& 2(\cos k_{x}+\cos k_{y}+\cos k_{z}) t_{m},\\
V_{\text{s}m}(\boldsymbol{k}) &=& 2i\sin k_{m} v_{\text{s}m},
\end{eqnarray}
where we took the lattice constant as unity.
\textcolor{black}{We take $t_{x}=t_{y}=t_{z}\equiv t_{p}$ and $v_{sx}=v_{sy}=v_{sz}$ that gives $V_{\text{s}x}=V_{\text{s}y}= V_{\text{s}z} \equiv iV/\sqrt{3}$ in accord with the cubic symmetry of the lattice.}
}

{

The atomic limit of the model given by ${\cal H}$ corresponds to the case of vanishing s-p hybridization \textcolor{black}{by taking} $V_{\text{s}m}=0$, i.e. when the  sites are decoupled. In this limit, the p state splits into \textcolor{black}{a twofold ($j=1/2$) and a fourfold ($j=3/2$) degenerate multiplet} with \textcolor{black}{energy} $-\lambda$ and $\lambda/2$, respectively, which gives the SO splitting $\Delta_\text{p}=3/2\lambda$ {\color{black}in agreement with Eq.~\eqref{HydrogenLikeAtomSOCSplitting}}.
}

{\color{black} The Elliott-Yafet theory involves the relevant SO matrix elements between adjacent s and p states that are mixed due to the presence of hybridization. We note that the matrix element \emph{vanishes} without hybridization, i.e. for the atomic limit. We obtain the spin admixed states due to SOC and the SO matrix elements $L = L_\text{s;p}$ by diagonalizing the Hamiltonian, ${\cal H}_{0}$, and by applying first-order perturbation theory with respect to the SO interaction. The details of the calculation are given in the Supplementary Material.
}

{
Figure~\ref{fig-energy}. shows the effect of non-zero hybridization on the originally pure s and p states in Na.
Namely, the sixfold degenerate p state splits into a quartet $\{\widetilde{\text{p}};\alpha \sigma\}$ and a doublet due to the mixing with the above lying s state $\{\widetilde{\text{s}}; \sigma \}$.
{\color{black} Considering} the SOC as perturbation, it induces additional spin mixing as expressed in Eqs.~(\ref{eq-st1})-(\ref{eq-st2}).
For example, an originally spin-down state of the quartet with dominantly p-character becomes mixed with a spin-up (and spin-down as well) state of the doublet with dominantly s-character as
\begin{eqnarray}
| \widetilde{\text{p}}; a \widetilde{\downarrow} \rangle  &=& | \widetilde{\text{p}}; a \downarrow \rangle + \frac{{\cal L}}{\Delta E} \left( \frac{1}{\sqrt{2}} | \widetilde{\text{s}}; \uparrow \rangle -\frac{i}{\sqrt{2}} | \widetilde{\text{s}}; \downarrow \rangle     \right), \label{eq-st1n}
\end{eqnarray}
\noindent where ${\cal L}$ is the magnitude of the SO matrix element between the quartet and the doublet and it reads
\begin{eqnarray}
{\cal L} &=&  \frac{\lambda V}{\sqrt{4V^2 + \left(\widetilde{E} +  \sqrt{\widetilde{E}^2+4V^2} \right)^2}}\label{eq-matre}
\end{eqnarray}
with $\widetilde{E} =  E_\text{s} - E_\text{p} + (\varepsilon_\text{s}(\boldsymbol{k}) -\varepsilon_\text{p}(\boldsymbol{k}))$. $\Delta E$ is the energy difference between the two states given as
\begin{eqnarray}
\Delta E = \frac{1}{2} \left( \widetilde{E} + \sqrt{ \widetilde{E}^2+4V^2}\right).
\end{eqnarray}
}

\begin{figure}
\centering
\includegraphics[width=0.98\hsize]{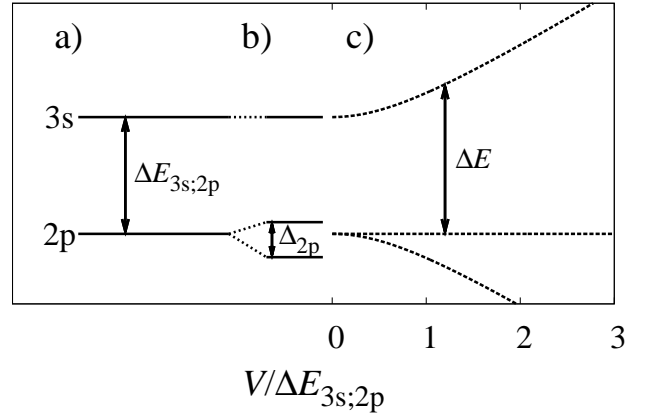}
\caption{Level splitting of 3s and 2p states (not to scale) in Na: a) without SOC and without hybridization; b) in the atomic limit, i.e. under vanishing hybridization $V=0$; and c) under non-zero s-p hybridization, $V$, without SOC.}
\label{fig-energy}
\end{figure}

{
By summing up the relevant Elliott-Yafet contributions (for details, see the Supplementary Material), we obtain the spin-orbit admixture coefficient $b$ as
\begin{eqnarray}
b =  \frac{L}{\Delta E}   = \frac{2 {\cal L}}{\Delta E}.
\end{eqnarray}
In the atomic limit ($V=0$), the SO matrix element vanishes between the s and p states as expected.
However, ${\cal L}$ and the corresponding SOAC are determined by the atomic energy splitting $E_\text{s}-E_\text{p}=\Delta E_\text{s;p}$,  the band parameter $\varepsilon_\text{s}(\boldsymbol{k}) -\varepsilon_\text{p}(\boldsymbol{k})$, the SO interaction $\lambda$, and the hybridization $V$.
}

{
Now, we turn to study the ratio $b/b_{\rm MB}$ of the spin-orbit admixture coefficients, where the Monod-Beuneu estimation, $b_{\text{MB}}$, of {\color{black}the spin-orbit admixture parameter} is given as
\begin{eqnarray}
b_{\text{MB}} &\equiv& \frac{\Delta_\text{p}}{\Delta E_\text{s;p}}
\end{eqnarray}
since the SO matrix element is approximated by the atomic SO energy splitting $\Delta_\text{p}$ for the p orbital in their picture.
The limit of $[\varepsilon_\text{s}(\boldsymbol{k}) -\varepsilon_\text{p}(\boldsymbol{k})]/\Delta E_\text{s;p} \rightarrow 0$ corresponds to the case where the bandwidths \textcolor{black}{given as $4t_{s}$ and $4t_{p}$ for the s and p bands, respectively,} are assumed to be much smaller than the s-p energy separation $\Delta E_\text{s;p}$.
By taking $\varepsilon_\text{s}(\boldsymbol{k}) -\varepsilon_\text{p}(\boldsymbol{k})=0$ and fixing the SOC interaction strength, $\lambda$, from the atomic energy splitting $\Delta_\text{p}$  as it is given in Eq.~(\ref{HydrogenLikeAtomSOCSplitting}), the ratio $b/b_{\rm MB}$ becomes a universal function of $V/\Delta E_\text{s;p}$ which is shown in the upper panel of Fig.~\ref{fig-SOAC} (the details are given in the Supplementary Material where the case of non-zero band parameter \textcolor{black}{$\varepsilon_\text{s}(\boldsymbol{k}) -\varepsilon_\text{p}(\boldsymbol{k})$, i.e. allowing finite bandwidths,} is also discussed).
}

{
Next we take the atomic values of $\Delta E_\text{s;p}$ for the alkali metals Na, K, Rb and Cs from Ref.~\onlinecite{yafet1963g} and estimate the hybridization parameter as
\begin{eqnarray}
V \equiv  \frac{c_{V}}{d},
\end{eqnarray}
where $d$ is the lattice constant being typically 4-6\, \AA, \textcolor{black}{and $c_{V}$ is a constant}.
The lower panel of Fig.~\ref{fig-SOAC} shows the ratio $b/b_{\rm MB}$ calculated for the different
alkali metals as a function of the hybridization coefficient $c_{V}$.
}

\begin{figure}
\centering
\subfigure{\includegraphics[width=0.9\hsize]{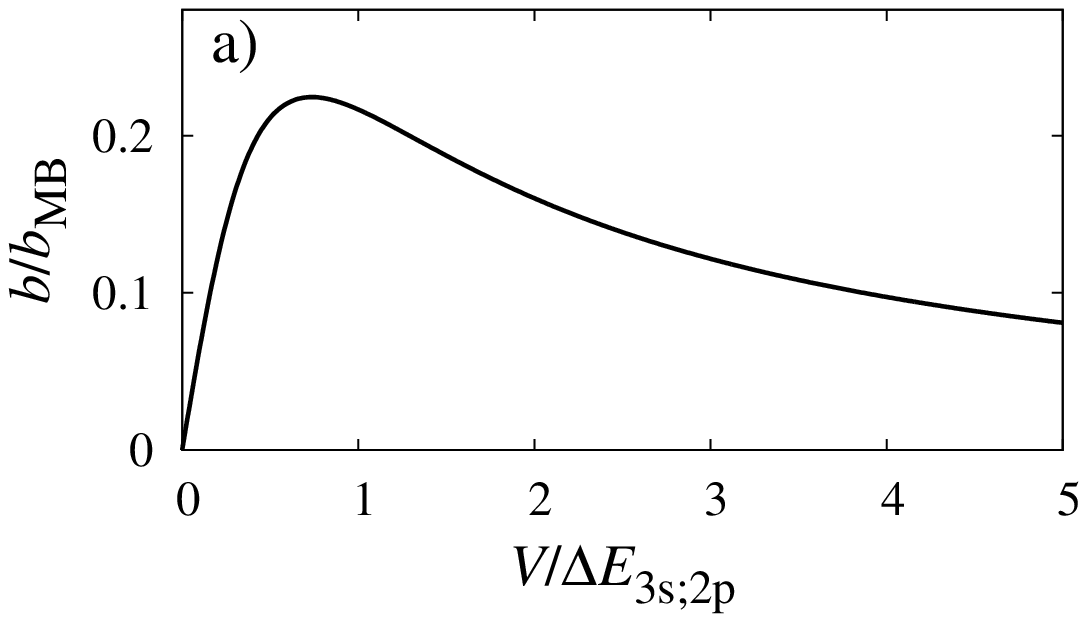}} \\
\subfigure{\includegraphics[width=0.9\hsize]{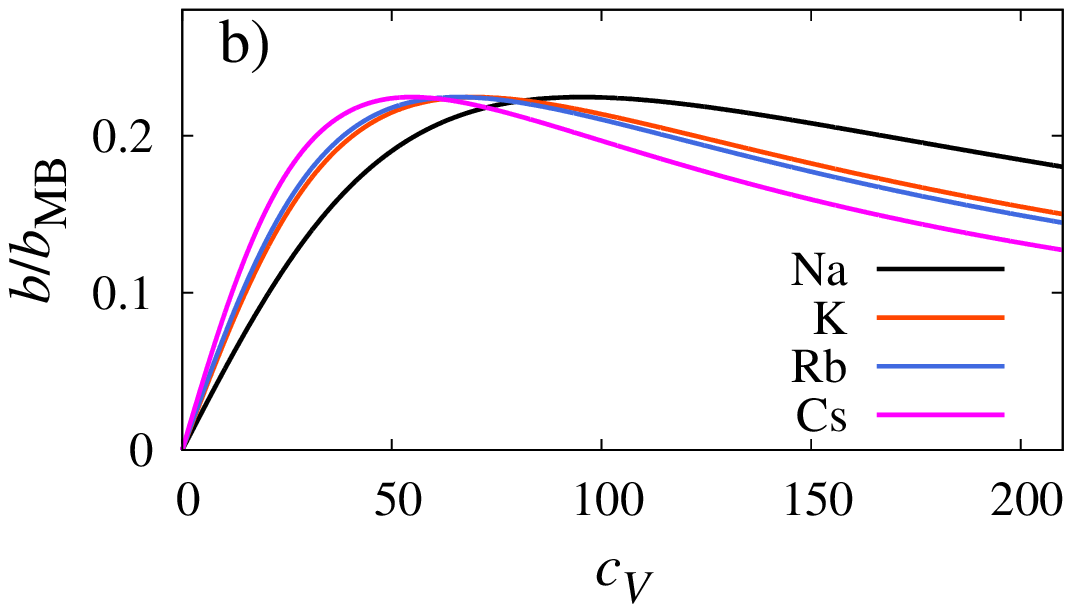}}

\caption{a) The calculated ratio, $b/b_{\rm MB}$, as a function of the s-p hybridization $V/\Delta E_\text{s;p}$ with $\varepsilon_\text{s}(\boldsymbol{k}) -\varepsilon_\text{p}(\boldsymbol{k})=0$, b) the calculated ratio, $b/b_{\rm MB}$, for the {\color{black}various} alkali metals as a function of the hybridization coefficient, $c_{V}$ (the atomic parameter values are taken from Ref.~\onlinecite{yafet1963g}).
}
\label{fig-SOAC}
\end{figure}

{
We observe that the calculated SOAC markedly differs from the Monod-Beuneu estimation \textcolor{black}{in the entire range of the hybridization used in the calculation}.
Reasons for the discrepancy can be that i) we estimate the SO interaction strength $\lambda$ from the atomic energy splitting $\Delta_\text{p}$ of the p orbital, which might give smaller $\lambda$ and therefore smaller SOAC than the real ones; ii) our model is too simple: although it \textcolor{black}{yields} non-zero SO matrix element between the adjacent s and p states, the only tunable parameter is the hybridization, $V$, if we assume small bandwidths.
Nevertheless, based on the evaluation of the s-p hybridization parameter as $V_{\text{sp}} \sim 4.2$\,eV in graphene \cite{HernandoPRB2006}, we estimate the hybridization coefficient $c_{V}$ being in the range of $1-40$\, eV$\cdot$\AA\, that gives the hybridization $V$ as $0.1-10$\,eV.
\textcolor{black}{In this range, i.e. for $V/\Delta E_\text{s;p} < 1$, the SOAC ratio depends linearly on $V/\Delta E_\text{s;p}$ as $b/b_{\text{MB}} \sim V/\Delta E_\text{s;p}$} (see the upper panel of Fig.~\ref{fig-SOAC}, and also Eq.~(B-4) of the Supplementary Material).
Assuming that the hybridization coefficient, $c_{V}$, does not change substantially among the alkali metals,
we obtain the following relations for the SOAC in the different alkali metals
\begin{eqnarray}
\left.\frac{b}{b_{\rm MB}}\right|_{\rm Cs} > \left.\frac{b}{b_{\rm MB}}\right|_{\rm Rb}, \,\,\,\,\,\,\,\,\,
\left.\frac{b}{b_{\rm MB}}\right|_{\rm Na, K} < \left.\frac{b}{b_{\rm MB}}\right|_{\rm Rb}\label{eq-relations}
\end{eqnarray}
from the lower panel of Fig.~\ref{fig-SOAC}.
Since the lattice constant does not vary much from Na to Cs either, the ratio $b/b_{\text{MB}}$ is roughly proportional to $1/\Delta E_\text{s;p}$, which explains the relations given in Eq.~(\ref{eq-relations}) because $\Delta E_\text{s;p}^{{\rm Cs}} < \Delta E_\text{s;p}^{{\rm Rb}}$ and $\Delta E_\text{s;p}^{{\rm Na,\, K}} > \Delta E_\text{s;p}^{{\rm Rb}}$ obtained from Ref.~\onlinecite{yafet1963g}.
}

{
\color{black}We compare the calculated result in Fig. \ref{fig-SOAC} and Eq.~(\ref{eq-relations})., with the empirical result in Fig.~\ref{lperd} and Table~\ref{tab:lambdas}. We find that our model does not reproduce the empirical ratios of $b/b_{\rm MB}$ \textcolor{black}{quantitatively}, however the tendency of the ratios for the different alkali metals are in fact accurately reproduced.
}

{
Although our model cannot provide a comprehensive description for even the simple alkali metals, it conveys the message that the real SO matrix elements, and therefore spin-relaxation mechanisms, depend on the nature of band structure and also on microscopic details such as the mixing of the s and p orbitals {\color{black}and that by no means can the atomic spin-orbit coupling be used directly to calculate the spin-relaxation properties in metals.}
For real systems, first principles calculations are required which could account for the exact matrix elements and the corresponding spin-orbit admixture coefficients.
}

\section{Conclusions}

We revisited the seminal contribution of Monod and Beuneu, who scaled the experimental ESR line-width data for elemental metals with the atomic spin-orbit coupling induced energy splitting and thus obtained a scaling with the electron momentum-scattering rate using a "universal" Gr\"{u}neisen-function. This approach is shown to be qualitative only and the proper description of the electron momentum-scattering calls for the inclusion of the Debye temperature and electron-phonon coupling, too. When this is considered, empirical spin-orbit admixture coefficients are obtained, which can serve as input for first principles calculations.

{
We provided a model calculation involving s and p states with spin-orbit coupling and we pointed out that in general the spin-orbit matrix elements present in the Elliott-Yafet theory are different from the SOC induced splitting of the atomic levels.
}

\section{Acknowledgements}
Enlightening discussions with A. J\'{a}nossy are gratefully acknowledged. Work supported by the ERC Grant Nr. ERC-259374-Sylo, by the Swiss National Science Foundation, by the Marie Curie Grant PIRG-GA-2010-276834, and the Hungarian Scientific Research Funds No. K106047.

\begin{widetext}

\section*{Supplementary information}
\appendix

This Supplementary Material is organized as follows: we first discuss the technical details of the calculations starting from the model Hamiltonian given in Eq.~(16) of the main text including the derivation of the relevant spin-orbit matrix elements and spin-orbit admixed states. Second, we extend the Elliott-Yafet formula given in Eq.~(3) of the main text to be appropriate to describe spin-relaxation in alkali metals within the Elliott-Yafet theory.

\section{Details of the calculations}

  \setcounter{equation}{0}  

The Fourier transform of the Hamiltonian given in Eq.~(16) reads as
\begin{eqnarray}
{\cal H}(\boldsymbol{k}) &=& {\cal H}_{0}(\boldsymbol{k}) + {\cal H}_{\text{SO}} (\boldsymbol{k});
\label{eq-hamm11}\\
{\cal H}_{0}(\boldsymbol{k}) &=&   \sum_{\boldsymbol{k}} \sum_{\sigma}  \sum_{m=\text{s},x,y,z} \varepsilon_{m}(\boldsymbol{k}) c^{\dag}_{\boldsymbol{k},m \sigma} c_{\boldsymbol{k}, m \sigma}
+ \sum_{\boldsymbol{k} }  \sum_{\sigma} \sum_{m=x,y,z}  \left( V_{\text{s}m}(\boldsymbol{k}) c^{\dag}_{\boldsymbol{k},\text{s} \sigma} c_{\boldsymbol{k}, m \sigma} + {\rm h.c.} \right)
+ E_\text{s}   \sum_{\sigma}   c^{\dag}_{\text{s} \sigma} c_{\text{s} \sigma} \nonumber\\
&+& E_{\text{p}}   \sum_{m=x,y,z}\sum_{\sigma}   c^{\dag}_{m \sigma} c_{m \sigma},
\label{eq-hamm12}\\
 {\cal H}_{\text{SO}} (\boldsymbol{k}) &=&   \lambda  \boldsymbol{L} \cdot \boldsymbol{S},\label{eq-hamm13}
\end{eqnarray}
where $c_{\alpha \sigma} = 1/\sqrt{N_{0}} \sum_{\boldsymbol{k}} c_{\boldsymbol{k}, \alpha \sigma}$ with $N_{0}$ being number of sites, and
\begin{eqnarray}
\varepsilon_{m}(\boldsymbol{k}) &=& 2(\cos k_{x}+\cos k_{y}+\cos k_{z}) t_{m},\\
V_{\text{s}m}(\boldsymbol{k}) &=& 2i\sin k_{m} v_{\text{s}m}.
\end{eqnarray}
The full Hamiltonain ${\cal H}(\boldsymbol{k})$ has the matrix form
\begin{eqnarray}
\hat{\cal H}(\boldsymbol{k}) &=&
\begin{pmatrix}
E_\text{s} + \varepsilon_\text{s}(\boldsymbol{k}) &  0 &  V_{\text{s}x}(\boldsymbol{k}) &  V_{\text{s}y}(\boldsymbol{k}) &  V_{\text{s}z}(\boldsymbol{k}) &  0 &  0 &  0 \\
0 &  E_\text{s} + \varepsilon_\text{s}(\boldsymbol{k})  &  0 &  0 &  0 &  V_{\text{s}x}(\boldsymbol{k}) &  V_{\text{s}y}(\boldsymbol{k})&  V_{\text{s}z}(\boldsymbol{k}) \\
V_{\text{s}x}(\boldsymbol{k})^{\ast} &  0 &  E_{\text{p}}+ \varepsilon_{x}(\boldsymbol{k})  &  - \frac{i}{2} \lambda &  0 &  0 &  0 &  \frac{1}{2} \lambda \\
V_{\text{s}y}(\boldsymbol{k})^{\ast} &  0 &  \frac{i}{2} \lambda &  E_{\text{p}}+\varepsilon_{y}(\boldsymbol{k}) &  0 &  0 &  0 &  -\frac{i}{2} \lambda \\
V_{\text{s}z}(\boldsymbol{k})^{\ast} &  0 &  0 &  0 & E_{\text{p}}+ \varepsilon_{z}(\boldsymbol{k}) &  -\frac{1}{2} \lambda &  \frac{i}{2} \lambda &  0 \\
0 &  V_{\text{s}x}(\boldsymbol{k})^{\ast} &  0 &  0 &  -\frac{1}{2} \lambda &  E_{\text{p}}+\varepsilon_{x}(\boldsymbol{k}) &  \frac{i}{2} \lambda &  0 \\
0 &  V_{\text{s}y}(\boldsymbol{k})^{\ast} &  0 &  0 &  -\frac{i}{2} \lambda &  -\frac{i}{2} \lambda & E_{\text{p}}+ \varepsilon_{y}(\boldsymbol{k}) &  0 \\
0 &  V_{\text{s}z}(\boldsymbol{k})^{\ast} &  \frac{1}{2} \lambda &  \frac{i}{2} \lambda & 0 & 0 &  0 &  E_{\text{p}}+\varepsilon_{z}(\boldsymbol{k})
\end{pmatrix} \nonumber\\
\label{eq-ham3}
\end{eqnarray}
writing in the basis $\left[ |\text{s}\uparrow;\boldsymbol{k} \rangle, |\text{s}\downarrow;\boldsymbol{k} \rangle, |\text{p}_{x}\uparrow;\boldsymbol{k} \rangle, |\text{p}_{y}\uparrow;\boldsymbol{k} \rangle ,|\text{p}_{z}\uparrow;\boldsymbol{k} \rangle, |\text{p}_{x}\downarrow;\boldsymbol{k} \rangle, |\text{p}_{y}\downarrow;\boldsymbol{k} \rangle, |\text{p}_{z}\downarrow;\boldsymbol{k} \rangle \right]$.
In the following, we will omit to write explicitly the $\boldsymbol{k}$-dependence in the expressions of the states.

In the presence of non-zero hybridization the originally six-fold degenerate p state splits into a quartet $\{\widetilde{\text{p}}; \alpha \sigma\}$ and a doublet due to the s-p mixing with the originally s-symmetric doublet state $\{\widetilde{\text{s}}; \sigma \}$ as it is shown in Fig.~5 of the main text.
By diagonalizing the Hamiltonian $\hat{\cal H}_{0}(\boldsymbol{k})$, we obtain the states of the quartet $\{\widetilde{\text{p}};\alpha \sigma\}$ and the doublet $\{\widetilde{\text{s}}; \sigma \}$ as
\begin{eqnarray}
| \widetilde{\text{p}}; {a} \downarrow \rangle &=& \frac{1}{\sqrt{6}} \left( -2|\text{p}_{x} \downarrow \rangle + |\text{p}_{y} \downarrow \rangle + |\text{p}_{z} \downarrow \rangle\right), \label{eq-q1}\\
| \widetilde{\text{p}}; {a} \uparrow \rangle &=& \frac{1}{\sqrt{6}} \left( -2|\text{p}_{x} \uparrow \rangle + |\text{p}_{y} \uparrow \rangle + |\text{p}_{z} \uparrow \rangle\right),\label{eq-q2}\\
| \widetilde{\text{p}}; {b} \downarrow \rangle &=& \frac{1}{\sqrt{2}} \left( - |\text{p}_{y} \downarrow \rangle + |\text{p}_{z} \downarrow \rangle\right),\label{eq-q3}\\
| \widetilde{\text{p}}; {b} \uparrow \rangle &=& \frac{1}{\sqrt{2}} \left( - |\text{p}_{y} \uparrow \rangle + |\text{p}_{z} \uparrow \rangle\right),\label{eq-q4}
\end{eqnarray}
and
\begin{eqnarray}
| \widetilde{\text{s}}; \downarrow \rangle &=& i\alpha | \text{s} \downarrow \rangle + \sqrt{\frac{1-\alpha^2}{3}} \left( |\text{p}_{x} \downarrow \rangle + |\text{p}_{y} \downarrow \rangle + |\text{p}_{z} \downarrow \rangle\right),\label{eq-s1}\\
| \widetilde{\text{s}}; \uparrow \rangle &=& i\alpha | \text{s} \uparrow \rangle + \sqrt{\frac{1-\alpha^2}{3}} \left( |\text{p}_{x} \uparrow \rangle + |\text{p}_{y} \uparrow \rangle + |\text{p}_{z} \uparrow \rangle\right)\label{eq-s2}
\end{eqnarray}
by assuming the following form for the hybridization \textcolor{black}{($\boldsymbol{k}\parallel(1,1,1)$)}: $V_{\text{s}x}(\boldsymbol{k}) =V_{\text{s}y}(\boldsymbol{k})= V_{\text{s}z}(\boldsymbol{k}) \equiv iV/\sqrt{3}$.
The coefficient $\alpha$ depends on the hybridization parameter $V$. The splitting between the above quartet and doublet is calculated as
\begin{eqnarray}
\Delta E = \frac{1}{2} \left( E_\text{s} - E_\text{p} + (\varepsilon_\text{s}(\boldsymbol{k}) -\varepsilon_\text{p}(\boldsymbol{k})) +  \sqrt{ [E_\text{s} - E_\text{p} + (\varepsilon_\text{s}(\boldsymbol{k}) -\varepsilon_\text{p}(\boldsymbol{k}))]^2+4V^2 }\right).
\end{eqnarray}

Switching on the SOC as perturbation, it induces additional spin mixing between the originally s- and p-symmetric states.
Since the SO interaction $ {\cal H}_{\text{SO}}$ does not have matrix element between the p-doublet and s-doublet, the spin mixing in first-order perturbation theory is determined by the SO matrix elements between the p-quartet and s-doublet given in Eqs.~(\ref{eq-q1})-(\ref{eq-q4}) and (\ref{eq-s1})-(\ref{eq-s2}), respectively, that are obtained as
\begin{eqnarray}
&&\langle \widetilde{\text{p}}; {a} \downarrow | {\cal H}_{\text{SO}} |  \widetilde{\text{s}}; \downarrow \rangle =
\langle \widetilde{\text{p}}; {a} \uparrow | {\cal H}_{\text{SO}} |  \widetilde{\text{s}}; \uparrow \rangle^{\ast} =  -\frac{i}{\sqrt{2}}{\cal L} , \\
&&\langle \widetilde{\text{p}}; {a} \downarrow | {\cal H}_{\text{SO}} |  \widetilde{\text{s}}; \uparrow \rangle =
- \langle \widetilde{\text{p}}; {a} \uparrow | {\cal H}_{\text{SO}} |  \widetilde{\text{s}}; \downarrow \rangle^{\ast} =  \frac{1}{\sqrt{2}}{\cal L} , \\
&&\langle \widetilde{\text{p}}; {b} \downarrow | {\cal H}_{\text{SO}} |  \widetilde{\text{s}}; \downarrow \rangle =
\langle \widetilde{\text{p}}; {b} \uparrow | {\cal H}_{\text{SO}} |  \widetilde{\text{s}}; \uparrow \rangle^{\ast} =  \frac{i}{\sqrt{6}}{\cal L} , \\
&&\langle \widetilde{\text{p}}; {b} \downarrow | {\cal H}_{\text{SO}} |  \widetilde{\text{s}}; \uparrow \rangle =
- \langle \widetilde{\text{p}}; {b} \uparrow | {\cal H}_{\text{SO}} |  \widetilde{\text{s}}; \downarrow \rangle^{\ast} =  \frac{(1+2i)}{\sqrt{6}}{\cal L},
\end{eqnarray}
where
\begin{eqnarray}
{\cal L} &=&  \frac{\lambda V}{\sqrt{4V^2 + \left(E_\text{s} - E_\text{p} + (\varepsilon_\text{s}(\boldsymbol{k}) -\varepsilon_\text{p}(\boldsymbol{k})) +  \sqrt{\left[E_\text{s} - E_\text{p} + (\varepsilon_\text{s}(\boldsymbol{k}) -\varepsilon_\text{p}(\boldsymbol{k})) \right]^2+4V^2} \right)^2}} = \frac{1}{2} \frac{\lambda V }{\sqrt{V^2 + \Delta E^2 }}.\label{eq-SOmatre}
\end{eqnarray}
\noindent

Then, the SOC induced spin admixed states of the p-quartet evolved from the s and p states are obtained as
\begin{eqnarray}
| \widetilde{\text{p}}; {a} \widetilde{\downarrow} \rangle  &=& | \widetilde{\text{p}}; {a} \downarrow \rangle + \frac{{\cal L}}{\Delta E} \left( \frac{1}{\sqrt{2}} | \widetilde{\text{s}}; \uparrow \rangle -\frac{i}{\sqrt{2}} | \widetilde{\text{s}}; \downarrow \rangle     \right), \label{eq-st1n}\\
| \widetilde{\text{p}}; {a} \widetilde{\uparrow} \rangle  &=& | \widetilde{\text{p}}; {a} \uparrow \rangle + \frac{{\cal L}}{\Delta E} \left( -\frac{1}{\sqrt{2}} | \widetilde{\text{s}} ;\downarrow \rangle + \frac{i}{\sqrt{2}} | \widetilde{\text{s}}; \uparrow \rangle     \right), \label{eq-st2n}\\
| \widetilde{\text{p}}; {b} \widetilde{\downarrow} \rangle  &=& | \widetilde{\text{p}}; {b} \downarrow \rangle + \frac{{\cal L}}{\Delta E} \left( \frac{(1+2i)}{\sqrt{6}} | \widetilde{\text{s}}; \uparrow \rangle + \frac{i}{\sqrt{6}} | \widetilde{\text{s}}; \downarrow \rangle     \right), \label{eq-st3n}\\
| \widetilde{\text{p}}; {b} \widetilde{\uparrow} \rangle  &=& | \widetilde{\text{p}}; {b} \uparrow \rangle + \frac{{\cal L}}{\Delta E} \left( \frac{(-1+2i)}{\sqrt{6}} | \widetilde{\text{s}} ;\downarrow \rangle - \frac{i}{\sqrt{6}} | \widetilde{\text{s}}; \uparrow \rangle     \right) \label{eq-st4n}
\end{eqnarray}
in the first order of the perturbation theory.

\section{Spin relaxation}

  \setcounter{equation}{0}  

\subsection{Formulation}

The central parameter in the Elliott-Yafet theory is the small coefficient ${\cal L}/\Delta E$ which describes the spin mixing of the adjacent states.
Since the Elliott-Yafet contributions are additive in Eq.~(3) of the main text as
\begin{eqnarray}
\frac{1}{\tau_{\text{s}}} &\sim & \sum_{n= \widetilde{\text{p}};{a} \downarrow,\widetilde{\text{p}};{a} \uparrow, \widetilde{\text{p}};{b} \downarrow, \widetilde{\text{p}};{b} \uparrow}  \sum_{m=\widetilde{\text{s}}; \downarrow, \widetilde{\text{s}};  \uparrow}
\frac{ |\langle n | {\cal H}_{\text{SO}} | m \rangle|^2}{\Delta E^2}  \frac{1}{\tau} =  4 \left( \frac{{\cal L}}{\Delta E}\right)^2 \frac{1}{\tau} \equiv  \left( \frac{L}{\Delta E}\right)^2 \frac{1}{\tau},
\label{elliottrel1n}
\end{eqnarray}
we define the "total" SO matrix element $L$ between the originally s- and p-symmetric states as
\begin{eqnarray}
L \equiv 2 {\cal L},
\end{eqnarray}
which gives the spin-orbit admixture parameter $b$ as
\begin{eqnarray}
b =  \frac{L}{\Delta E}   = \frac{2 {\cal L}}{\Delta E}.\label{eq-soac1}
\end{eqnarray}

\begin{figure}
\centering
\includegraphics[width=0.48\hsize]{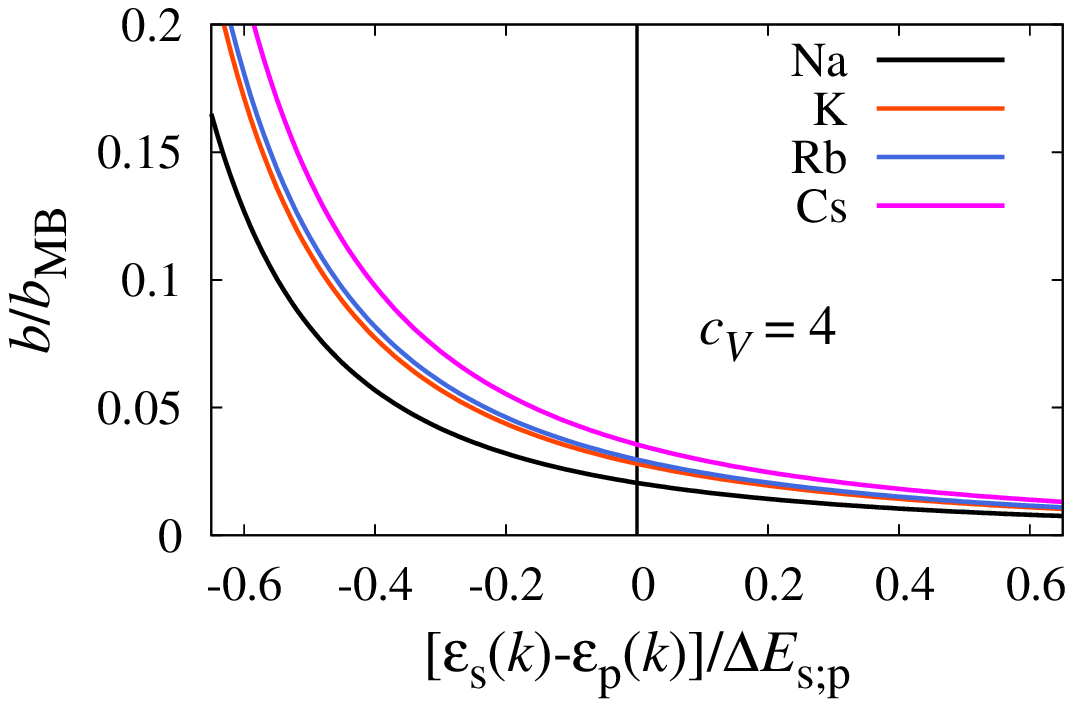}
\includegraphics[width=0.48\hsize]{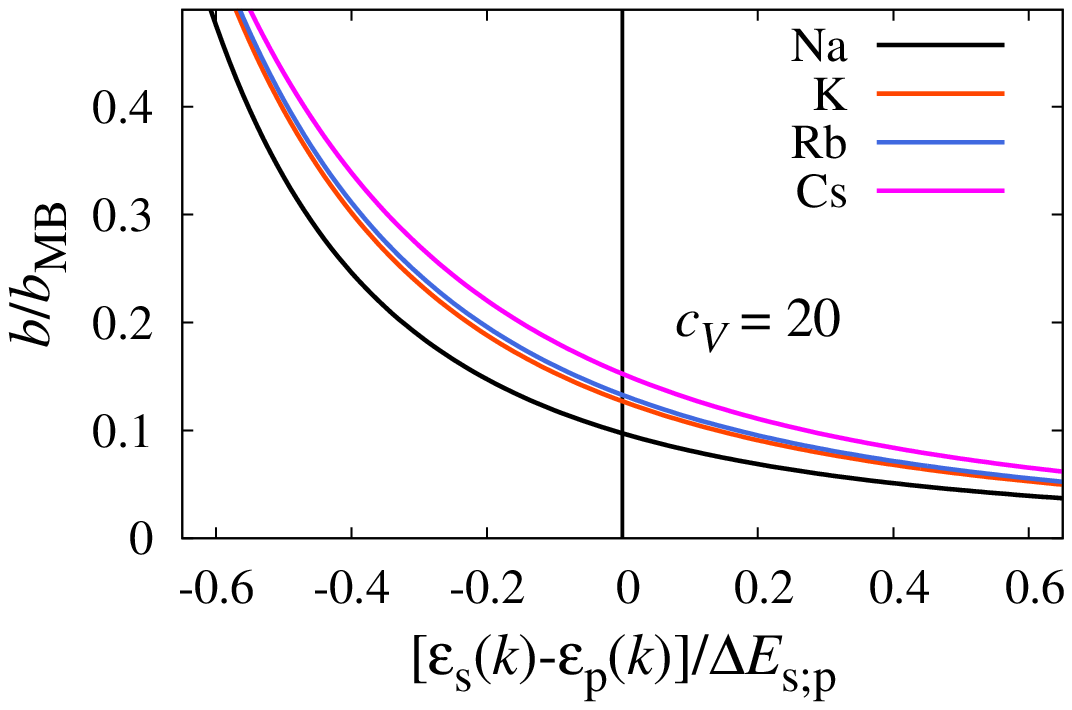}
\caption{The calculated ratio $b/b_{\rm MB}$ for alkali metals as a function of $[\varepsilon_\text{s}(\boldsymbol{k}) -\varepsilon_{\text{p}}(\boldsymbol{k})]/\Delta E_\text{s;p}$ with two different hybridization coefficients $c_{V}=4$ and $c_{V}=20$\,eV$\cdot$\AA. The atomic parameter values are taken from Ref.~[\onlinecite{yafet}].}
\label{fig-SOAC}
\end{figure}

\subsection{Spin-orbit admixture coefficient in alkali metals}

Using Eqs.~(\ref{eq-SOmatre}), (\ref{eq-soac1}), and the estimation $\lambda=2/3\Delta_{\text{p}}$ for the SO interaction strength, the ratio $b/b_{\rm MB}$ is expressed as
\begin{eqnarray}
\frac{b}{b_{\rm MB}} &=& \frac{4 \sqrt{2}}{3}  \left(\frac{V}{\Delta E_\text{s;p}}\right)
\frac{1}{\left( 1+\alpha/\Delta E_\text{s;p} + \sqrt{(1+\alpha/\Delta E_\text{s;p})^2 + 4 (V/\Delta E_\text{s;p})^2} \right)} \nonumber\\
&\times&
\frac{1}{\sqrt{  (1+\alpha/\Delta E_\text{s;p})^2 + 4 (V/\Delta E_\text{s;p})^2 +  (1+\alpha/\Delta E_\text{s;p})\sqrt{(1+\alpha/\Delta E_\text{s;p})^2 + 4 (V/\Delta E_\text{s;p})^2}  }},
\label{eq-soac2}
\end{eqnarray}
where $\Delta E_\text{s;p} = E_\text{s}-E_\text{p}$,  $b_{\text{MB}}$ is the Monod-Beuneu estimation as $ b_{\text{MB}}= \Delta_{\text{p}}/\Delta E_\text{s;p}$,
and $\alpha = (\varepsilon_\text{s}(\boldsymbol{k}) -\varepsilon_\text{p}(\boldsymbol{k}))$.

The band parameter $\alpha$ is related to the s and p bandwidths $W_\text{s}$ and $W_{\text{p}}$ since we may associate $W_{m}=\varepsilon_{m}(k=0)-\varepsilon_{m}(k=2\pi)=4t_{m}$.
The limit of $\alpha/\Delta E_\text{s;p} \rightarrow 0$ corresponds to the case where the bandwidths are assumed to be much smaller then the s-p energy separation $\Delta E_\text{s;p}$.
In this case, the ratio $b/b_{\text{MB}}$ given in Eq.~(\ref{eq-soac2}) depends only on $V/\Delta E_\text{s;p}$ leading a unique curve as a function of $V/\Delta E_\text{s;p}$ as it is shown in the upper panel of Fig.~5 in the main text.

Allowing non-zero value for $\alpha/\Delta E_\text{s;p}$ leads to separate curves for the different alkali metals.
We take the hybridization as $V=c_{V}/d$ with $d$ being the lattice constant, and fix the atomic energy splitting $\Delta E_\text{s;p}$ for Na, K, Rb and Cs from the literature\cite{yafet}.
In the main text we estimated the hybridization coefficient $c_{V}$ as being in the range of $1-40$\, eV$\cdot$\AA\, because it leads to hybridization with order of unity in eV as $0.1-10$\,eV.
Figure~\ref{fig-SOAC} shows the calculated ratio $b/b_{\rm MB}$ as a function of $[\varepsilon_\text{s}(\boldsymbol{k}) -\varepsilon_{\text{p}}(\boldsymbol{k})]/\Delta E_\text{s;p}$ for the different alkali metals in the regime where $(\varepsilon_\text{s}(\boldsymbol{k}) -\varepsilon_{\text{p}}(\boldsymbol{k})) \ll \Delta E_\text{s;p}$, where $\boldsymbol{k}$ should be taken as the Fermi wave vector $\boldsymbol{k}_{F}$.

\end{widetext}


\begin{thebibliography}{39}
\bibitem[{Not()}]{Note}
\bibinfo{note}{More precisely $L$ is the radial matrix element of $\xi(r)$.}
\end{thebibliography}
\end{document}